\documentclass[preprint,12pt]{elsarticle}

\usepackage{amssymb}
\usepackage{graphicx}
\usepackage{dcolumn}
\usepackage{bm}
\usepackage{rotating}

\journal{Computer Physics Communications}

\newcommand{\kslash}{k\kern-1ex /}
\newcommand{\pslash}{p\kern-1ex /}
\newcommand{\qslash}{q\kern-1ex /}
\newcommand{\lslash}{l\kern-1ex /}
\newcommand{\sslash}{s\kern-1ex /}
\newcommand{\Dslash}{D\kern-1.2ex /}

\newcommand{\beqa}{\begin{eqnarray}}
\newcommand{\eeqa}{\end{eqnarray}}

\newcommand{\bd}{\begin{description}}
\newcommand{\ed}{\end{description}}

\newcommand{\ben}{\begin{eqnarray}}
\newcommand{\een}{\end{eqnarray}}
\newcommand{\nn}{\nonumber}
\def\lsim{\raise0.3ex\hbox{$<$\kern-0.75em\raise-1.1ex\hbox{$\sim$}}}
\def\gsim{\raise0.3ex\hbox{$>$\kern-0.75em\raise-1.1ex\hbox{$\sim$}}}
\def\simgt{\rlap{\lower 6.0 pt\hbox{$\mathchar \sim$}}\raise 2.5pt \hbox {$>$}}
\def\simlt{\rlap{\lower 6.0 pt\hbox{$\mathchar \sim$}}\raise 2.5pt \hbox {$<$}}

\newcommand{\csw}{{c_{\rm SW}}}

\newcommand{\low}[1]{\smash{\lower1.65ex\hbox{#1}}}

\begin{document}

\begin{frontmatter}



\title{Application of preconditioned block BiCGGR to 
the Wilson-Dirac equation
with multiple right-hand sides in lattice QCD}


\author{H.~Tadano$^{a,b}$, Y.~Kuramashi$^{c,b}$, T.~Sakurai$^{a}$}

\address{
 $^a$Department of Computer Science, University of Tsukuba,\\ Tsukuba, Ibaraki 305-8573, Japan\\
 $^b$Center for Computational Sciences, University of Tsukuba,\\ Tsukuba, Ibaraki 305-8577, Japan\\
 $^c$Graduate School of Pure and Applied Sciences, University of Tsukuba,\\ Tsukuba, Ibaraki 305-8571, Japan\\
}

\begin{abstract}
There exist two major problems in 
application of the conventional block BiCGSTAB method to
the $O(a)$-improved Wilson-Dirac equation with multiple
right-hand-sides: One is the deviation 
between the true and the recursive residuals. The other is
the convergence failure observed at smaller quark masses for enlarged
number of the right-hand-sides.
The block BiCGGR algorithm which was recently proposed by the authors
succeeds in solving the former problem.
In this article we show that a preconditioning technique allows us to
improve the convergence behavior for increasing number 
of the right-hand-sides. 
\end{abstract}


\end{frontmatter}


\thispagestyle{plain}

\section{Introduction}
\label{sec:intro}

This paper is the third in a series of 
publications\cite{blockbicggr_th,blockbicggr_app}
on a new block Krylov subspace method called block BiCGGR.
In Ref.~\cite{blockbicggr_th} we proposed the algorithm 
which successfully removes the deviation between the 
true and the recursive residuals found in the block BiCGSTAB method.
Reference~\cite{blockbicggr_app} is devoted to the application 
of the new algorithm to solving the $O(a)$-improved Wilson-Dirac
equations in lattice QCD.  Although the significant cost
reduction is achieved by both the algorithmic efficiency and the
cache-aware implementation technique, there remains one concern that
the increase of $L$, which denotes the number of the right-hand sides,  
makes the convergence of the algorithm at lighter quark masses difficult. 
 
In this paper we investigate the effects of a preconditioning on the
convergence properties of the block BiCGGR method 
in solving the $O(a)$-improved Wilson-Dirac
equations in lattice QCD. 
For a comparative purpose we employ the same gauge configurations  
as in Ref.~\cite{blockbicggr_app}.
We focus on the lightest quark mass
used in Ref.~\cite{blockbicggr_app}, which was the most difficult case
to attain the convergence with the block BiCGGR method.
As a preconditioner we incorporate the inner solver with the Jacobi
method. The convergence behavior is examined 
by varying the iteration number $j$ of the Jacobi method.
We observe that the convergence properties are improved by the
preconditioner so that the block BiCGGR method retains its efficiency for
wider range of $L$. 
For $j\ge 12$ 
the computational cost with $L=12$ is reduced down to 10\% of that with $L=1$
showing stabilized convergence behaviors.

This paper is organized as follows. In Sec.~\ref{sec:algorithm} we explain
the algorithmic details of the block BiCGGR with the 
inner solver of the Jacobi method. We present the results of the
numerical tests in Sec.~\ref{sec:test}. Conclusions and discussions are
summarized in Sec.~\ref{sec:conclusion}. 

\section{Preconditioned block BiCGGR}
\label{sec:algorithm}
  
We consider to solve the linear systems with the
multiple right-hand sides expressed as
\ben
AX=B,
\label{eq:linear_mrhs}
\een
where $A$ is an $N\times N$ complex sparse non-Hermitian matrix.
$X$ and $B$ are $N\times L$ complex rectangular matrices given by
\ben
X&=&\left(\bm{x}^{(1)},\dots,\bm{x}^{(i)},\dots,\bm{x}^{(L)} \right),\\
B&=&\left(\bm{b}^{(1)},\dots,\bm{b}^{(i)},\dots,\bm{b}^{(L)} \right).
\een
In the case of the Wilson-Dirac equation the matrix dimension is
given by
$N=L_x\times L_y\times L_z\times L_t \times 3 \times 4$
with $L_x\times L_y\times L_z\times L_t$ the volume of 
a hypercubic four-dimensional lattice.
$L$ is the number of the right-hand-side vectors which is called
the source vectors in lattice QCD.
Throughout this paper the specific matrix structure of the $O(a)$-improved 
Wilson-Dirac equation is not necessary. 
We refer the readers who may be interested in it to 
Sec. 2 of Ref.~\cite{blockbicggr_app}.

The details of the block BiCGGR algorithm are presented 
in Refs.~\cite{blockbicggr_th,blockbicggr_app}.
The preconditioned block BiCGGR method 
with $M$ an $N \times N$ preconditioning matrix such that $M \approx A^{-1}$
is described as follows:

\begin{center}
\begingroup
\renewcommand{\arraystretch}{1.2}
\begin{tabular}{l}
 $X_0 \in \mathbb{C}^{N \times L}$ is an initial guess, \\
 {\bf Compute} $R_0 = B - AX_0$, \\
 {\bf Set} $P_0 = R_0$, \\
 {\bf Choose} $\tilde{R}_0 \in \mathbb{C}^{N \times L}$, \\
 {\bf Preconditioning part:} $F_0 = MR_0$,\\
 {\bf Set} $V_0 = W_0 = AF_0$,\\
 {\bf For} $k=0,1,\dots, $ {\bf until} $\displaystyle{\max_i (\|\bm{r}_k^{(i)}\|_2 / \|\bm{b}^{(i)}\|_2 ) \leq \varepsilon }$ {\bf do:}\\
 \begingroup
 \setlength{\tabcolsep}{0.2em}
 \begin{tabular}{rcl}
  \multicolumn{3}{l}{{\bf Solve} $(\tilde{R}_0^{\mathrm{H}}V_k)\alpha_k = \tilde{R}_0^{\mathrm{H}}R_k$ for $\alpha_k$,} \\
  $\zeta_k$ & = & $\mathrm{Tr}[W_k^{\mathrm{H}}R_k] / \mathrm{Tr}[W_k^{\mathrm{H}}W_k]$,\\
  $S_k$ & = & $P_k - \zeta_k V_k$, \\
  $U_k$ & = & $S_k \alpha_k$, \\
  \multicolumn{3}{l}{{\bf Preconditioning part:} $G_k = MU_k$,}\\
  $Y_k$ & = & $AG_k$, \\
  $X_{k+1}$ & = & $X_k + \zeta_k F_k + G_k$, \\
  $R_{k+1}$ & = & $R_k - \zeta_k W_k - Y_k$, \\
  \multicolumn{3}{l}{{\bf Preconditioning part:} $F_{k+1} = MR_{k+1}$,}\\
  $W_{k+1}$ & = & $AF_{k+1}$, \\
  \multicolumn{3}{l}{{\bf Solve} $(\tilde{R}_0^{\mathrm{H}}R_k)\gamma_k = \tilde{R}_0^{\mathrm{H}}R_{k+1}/\zeta_k$ for $\gamma_k$,} \\
  $P_{k+1}$ & = & $R_{k+1} + U_k \gamma_k$, \\
  $V_{k+1}$ & = & $W_{k+1} + Y_k \gamma_k$, \\
 \end{tabular}
 \endgroup \\
 {\bf End for.}
\end{tabular}
\endgroup \\
\end{center}
In this paper we employ the Jacobi method as a preconditioner of the block BiCGGR method because of its practical parallelizability.
In this case the matrix $G_k = MU_k$ in the algorithm is calculated by the Jacobi method as follows:

\begingroup
\begin{displaymath}
 G_k = \left\{
 \renewcommand{\arraystretch}{1.3}
 \begin{array}{ll}
  \displaystyle{(I - A_{\mathrm{D}}^{-1}A)^jG_{k,0} + \sum_{i=0}^{j-1}\left(I - A_{\mathrm D}^{-1}A \right)^i A_{\mathrm D}^{-1}U_k}, & j \geq 1, \\
  \displaystyle{U_k}, & j = 0,
 \end{array}
 \right.
\end{displaymath}
\endgroup
where $j$, $G_{k,0}$, and $A_{\mathrm D}$ denote the number of iterations of the Jacobi method, the initial guess for the Jacobi method, 
and the diagonal part of the coefficient matrix $A$, respectively.
The preconditioning part of the block BiCGGR algorithm is computed by the matrix-vector multiplications.

The dominant part of the memory requirements, which is proportional to
$N$, is given by $16N(51+9L)$ Bytes without an additional contribution
from the preconditioner.
In a practical sense it would be sufficient that the effectiveness of the
preconditioner is retained up to $L\sim 10$, because
the memory requirements may become a constraint 
on the applicability of the block BiCGGR method once $L$ goes beyond 10.

\section{Numerical tests}
\label{sec:test}

\subsection{Choice of parameters}
\label{sec:param}

We employ the same quenched gauge configurations as in
Ref.~\cite{blockbicggr_app}, which are the statistically independent 10
samples generated with the Iwasaki gauge action at $\beta=2.575$ on a
$L_x\times L_y\times L_z\times L_t=16\times 16\times 16\times 32$ lattice. 
We choose one hopping parameter 
$\kappa=0.1359$ for the Wilson-Dirac equation with the
improvement coefficient $c_{\rm SW}=1.345$.  
The bare quark mass is defined by $m_q=(1/\kappa-1/\kappa_c)/2$
with $\kappa_c=0.136116(8)$. 
Note that this hopping parameter gives the lightest quark mass
in Ref.~\cite{blockbicggr_app} which was the most problematic case to
achieve the convergence with the block BiCGGR method for the fixed $L$. 
According to the results in Ref.~\cite{cppacs_nf2},
the physical pion mass is 221 MeV 
with $m_\pi/m_\rho=0.28$ at $\kappa=0.1359$.  
The lattice spacing is $a=0.1130$ fm determined by $m_\rho$. 

\subsection{Test environment}
\label{subsec:enviroment}

Numerical tests are performed on single node of a large-scale
cluster system called T2K-Tsukuba
which was also employed in the previous study\cite{blockbicggr_app}. 
The machine consists of 648 compute nodes providing
95.4Tflops of computing capability.
Each node consists of quad-socket, 2.3GHz Quad-Core AMD Opteron Model
8356 processors whose on-chip cache sizes are 64KBytes/core,
512KBytes/core, 2MB/chip for L1, L2, L3, respectively.  
Each processor has a direct connect memory interface to an 8GBytes 
DDR2-667 memory and three hypertransport links to connect other processors.
All the nodes in the system are connected through 
a full-bisectional fat-tree network 
consisting of four interconnection links of 8GBytes/sec aggregate
bandwidth with Infiniband.

\subsection{Results}
\label{sec:results}

\begin{figure*}[h]
\vspace{13mm}
\begin{center}
\includegraphics[width=110mm,angle=0]{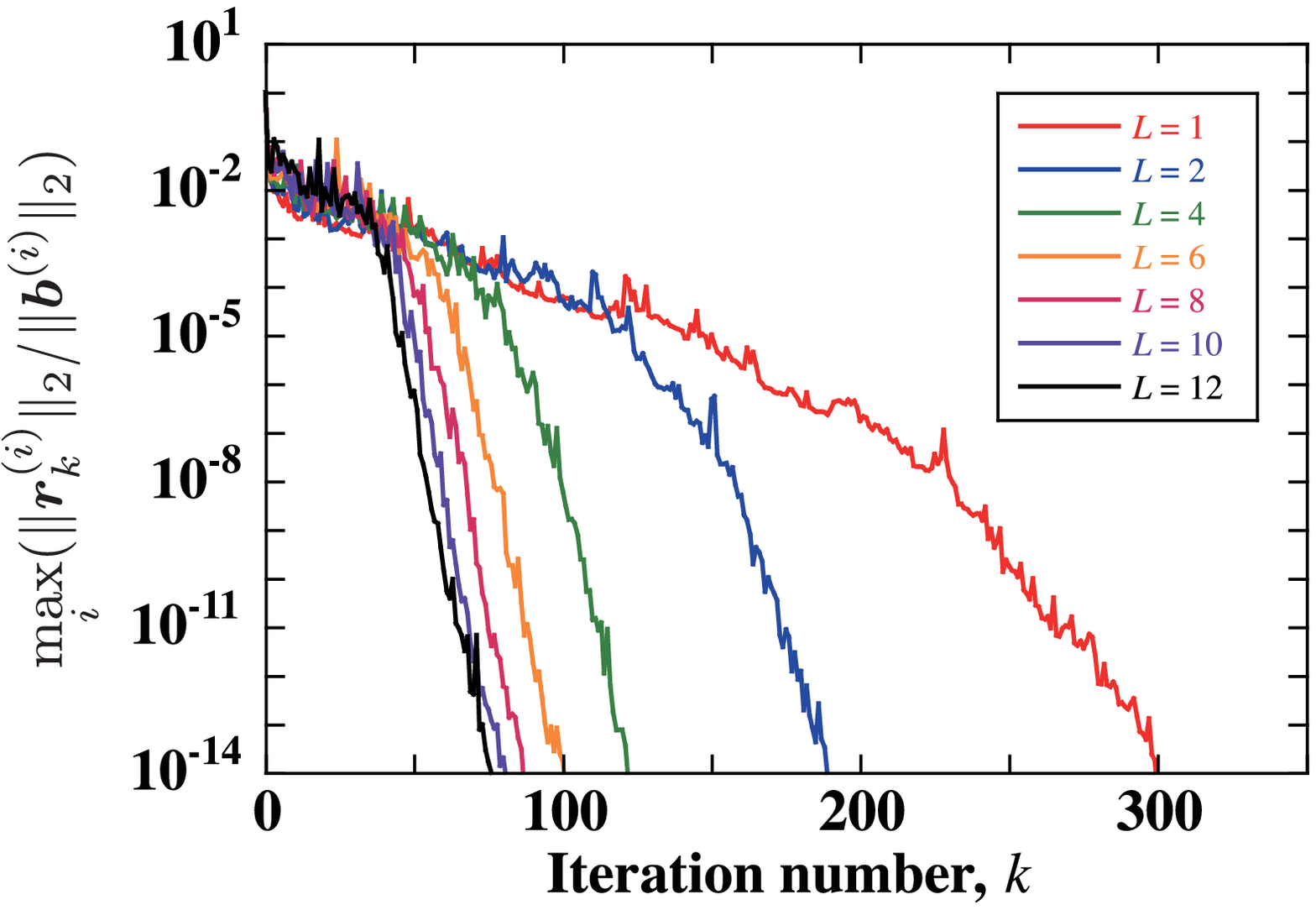}  
\end{center}
\vspace{-.5cm}
\caption{$L$ dependence of the convergence behaviors 
for the $j=12$ case. All the measurements are performed on
the same configuration.}
\label{fig:conv_j12}
\end{figure*}

\begin{figure*}[h]
\vspace{13mm}
\begin{center}
\includegraphics[width=110mm,angle=0]{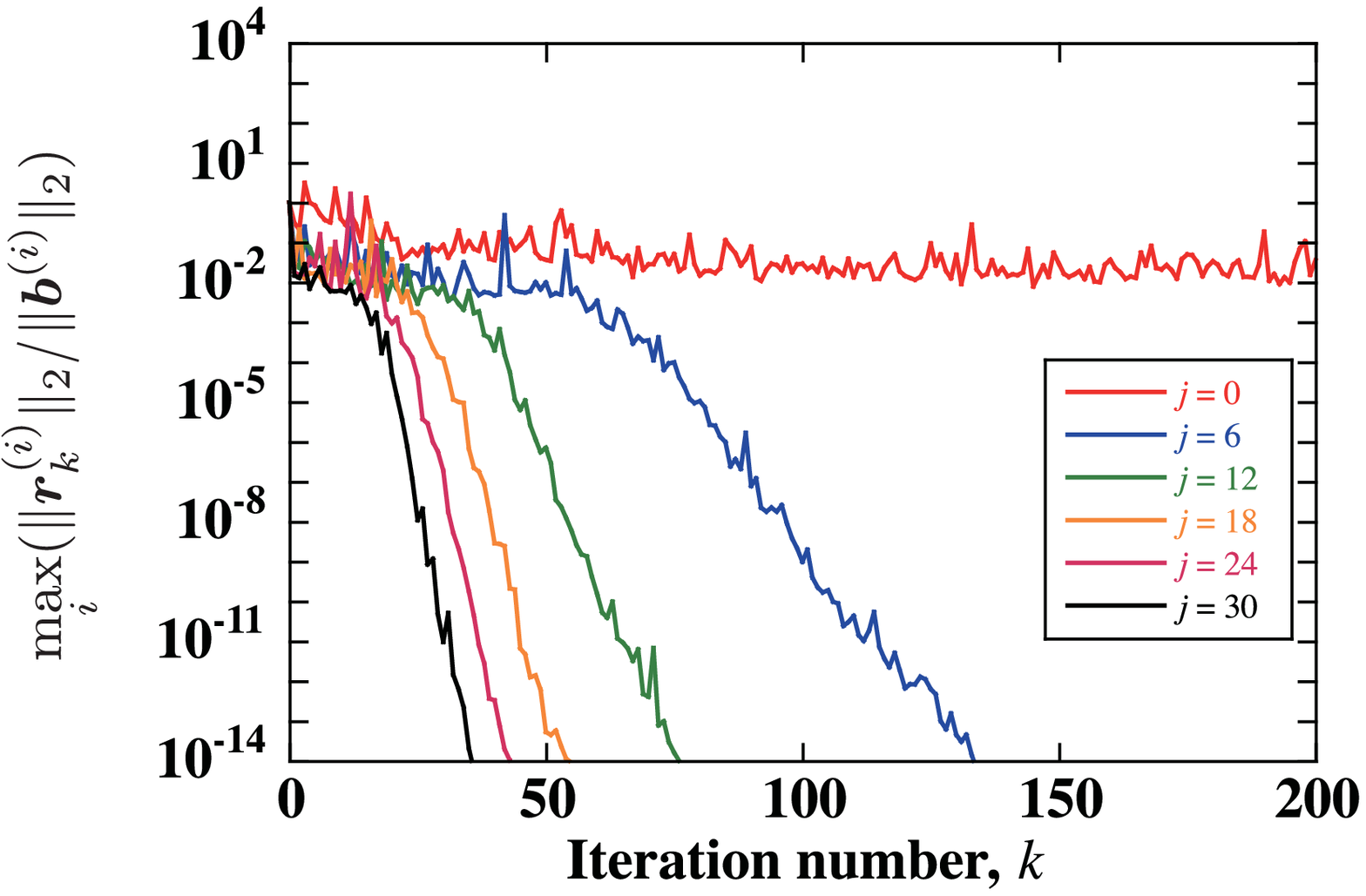}  
\end{center}
\vspace{-.5cm}
\caption{$j$ dependence of the convergence behaviors 
for the $L=12$ case. All the measurements are performed on
the same configuration.}
\label{fig:conv_L12}
\end{figure*}

In Table~\ref{tab:iteration} we list the outer iteration number to solve
the Wilson-Dirac equation with the  
preconditioned block BiCGGR algorithm
as a function of $L$ and the inner iteration number $j$.
The initial guess for the block BiCGGR method and the Jacobi method 
is set to the zero matrix.
The matrix $\tilde{R}_0$ for the block BiCGGR method is chosen as $R_0$.
We employ rather stringent tolerance of
$\max_i(\Vert \bm{r}_k^{(i)}\Vert_2/\Vert \bm{b}^{(i)}\Vert_2) \le
10^{-14}$ with $\bm{r}_k^{(i)}$ the recursive residual in the 
$k$-th outer iteration and  $\bm{b}^{(i)}$ a unit vector whose $i$-th 
component is unity.
The results are averaged over 10 configuration samples. 
In some combinations of $L$ and $j$ we find the convergence failure:
The residual ceases decreasing and starts to increase gradually
without reaching the tolerance.
In this case we give the number of the configuration samples 
which show the convergence failure in Table~\ref{tab:iteration}. 
Note that the total number of the
matrix-vector multiplication denoted by \#MVM 
is given by the formula of $2[(j+1)k + 1]L$. 
This should be a more appropriate quantity to be compared. 
We give \#MVM/$L$ within the parentheses 
in each entry of Table~\ref{tab:iteration}.  
Most important point is that we are allowed to achieve 
the convergence for enlarged $L$ as the inner iteration number increases.
Secondly, \#MVM/$L$ can be reduced with an appropriate 
choice of $L$ and $j$. 
To illustrate the convergence behaviors we plot 
$\max_i(\Vert \bm{r}_k^{(i)}\Vert_2/\Vert \bm{b}^{(i)}\Vert_2)$ 
as a function of the outer iteration number $k$ choosing one 
configuration sample as a representative case. 
Figure~\ref{fig:conv_j12} shows the $L$ dependence with $j$ fixed at twelve.
We observe a characteristic feature that the convergence behaviors for
different $L$ are almost identical up to some iteration number, 
beyond which the convergence speed for larger $L$ is suddenly 
accelerated.
In Fig.~\ref{fig:conv_L12} we plot the $j$ dependence for the $L=12$
case. For $j=0$ the iteration is terminated when 
the residual of $\max_i(\Vert \bm{r}_k^{(i)}\Vert_2/\Vert
\bm{b}^{(i)}\Vert_2)$  reaches $10^2$ without achieving the convergence.
It may be surprising that both figures show a quite similar feature 
under the exchange of $j$ and $L$. 

In Table~\ref{tab:time} we present the execution time divided by $L$ as
a function of $L$ and $j$. 
A remarkable cost reduction is observed. The best case is the
combination of $L=12$ and $j=12$ where the cost is just 10\% 
of that for the unimproved case with $L=1$ and $j=0$. 
In a practical use it is reasonable to choose 
$L=12$ with $j\ge 12$ as default parameters:
We observe that the stabilized convergence properties with
less execution time. If the convergence is failed by some possibility,
you just repeat the inversion with smaller $L$.

\begin{figure*}[h]
\vspace{13mm}
\begin{center}
\includegraphics[width=110mm,angle=0]{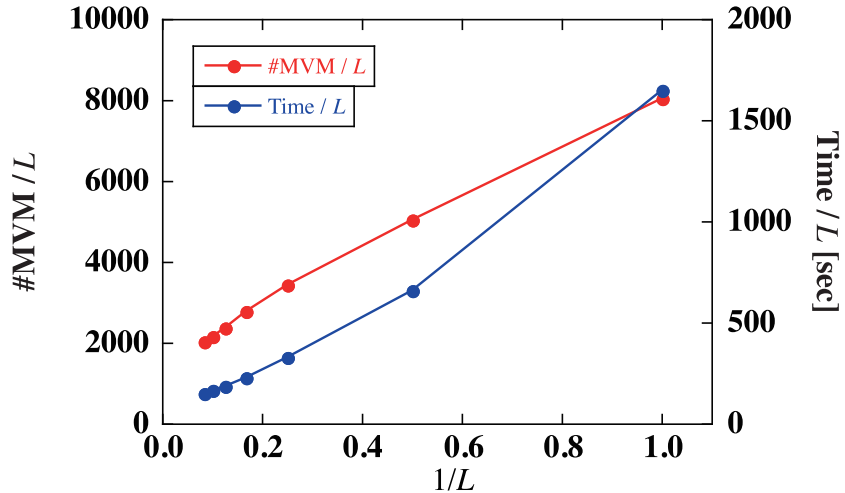}  
\end{center}
\vspace{-.5cm}
\caption{$L$ dependence of \#MVM/$L$ (red) and
the execution time divided by $L$ (blue) for the preconditioned 
block BiCGGR method with $j=12$. All the
 results are averaged over 10 configuration samples.}
\label{fig:cost}
\end{figure*}

There are two key ingredients for this remarkable
achievement. One is the algorithmic improvements thanks to the block
BiCGGR: For the given value of $j$, 
\#MVM/$L$ monotonically decreases as a function of $L$.
The other is the efficiency of the cache-aware implementation technique 
for multiple $L$.
The situations are depicted in Fig.~\ref{fig:cost} 
with the choice of $j=12$.

\begin{table*}[htb]
\centering{
\caption{Outer iteration number as a function of $L$ and the inner 
iteration number $j$ to solve the Wilson-Dirac equation with the
preconditioned block BiCGGR method. Results are averaged over 10
 configuration samples. Fail means how many configuration samples
out of ten show the convergence failure.
\#MVM/$L$ is given in the parentheses. 
}
\label{tab:iteration}
\newcommand{\cc}[1]{\multicolumn{1}{c}{#1}}
   \begin{tabular}{r|r|r|r|r|r|r}
    \noalign{\hrule height 1.0pt}
    \multicolumn{7}{c}{$\kappa = 0.1359$}\\
    \multicolumn{1}{c}{$L$} & \multicolumn{1}{c}{$j=0$} & \multicolumn{1}{c}{$j=6$} & \multicolumn{1}{c}{$j=12$} & \multicolumn{1}{c}{$j=18$} & \multicolumn{1}{c}{$j=24$} & \multicolumn{1}{c}{$j=30$} \\
    \noalign{\hrule height 1.0pt}
     \low{$1$} & $   2437.4$ & $    536.4$ & $    309.4$ & $    224.4$ & $    176.4$ & $    145.8$ \\
                & ($   4875.8$) & ($   7510.6$) & ($   8045.4$) & ($   8528.2$) & ($   8821.0$) & ($   9040.6$) \\
    \hline
     \low{$2$} & $   1713.4$ & $    342.0$ & $    194.1$ & $    140.3$ & $    110.0$ & $     90.5$ \\
                & ($   3427.8$) & ($   4789.0$) & ($   5047.6$) & ($   5332.4$) & ($   5501.0$) & ($   5612.0$) \\
    \hline
     \low{$4$} & \low{Fail: $7$/$10$} & \low{Fail: $1$/$10$} & $    132.3$ & $     92.7$ & $     73.5$ & $     61.2$ \\
                &             &             & ($   3440.8$) & ($   3523.6$) & ($   3676.0$) & ($   3795.4$) \\
    \hline
     \low{$6$} & \low{Fail: $10$/$10$} & $    184.5$ & $    107.1$ & $     75.9$ & $     59.6$ & $     49.3$ \\
                &             & ($   2584.0$) & ($   2785.6$) & ($   2885.2$) & ($   2981.0$) & ($   3057.6$) \\
    \hline
     \low{$8$} & \low{Fail: $10$/$10$} & \low{Fail: $1$/$10$} & $     91.4$ & $     66.8$ & $     52.2$ & $     43.3$ \\
                &             &             & ($   2377.4$) & ($   2539.4$) & ($   2611.0$) & ($   2685.6$) \\
    \hline
     \low{$10$} & \low{Fail: $10$/$10$} & \low{Fail: $1$/$10$} & $     83.3$ & $     60.5$ & $     48.5$ & $     40.1$ \\
                &             &             & ($   2166.8$) & ($   2300.0$) & ($   2426.0$) & ($   2487.2$) \\
    \hline
     \low{$12$} & \low{Fail: $10$/$10$} & $    144.2$ & $     78.3$ & $     57.5$ & $     46.4$ & \low{Fail: $1$/$10$} \\
                &             & ($   2019.8$) & ($   2036.8$) & ($   2186.0$) & ($   2321.0$) &             \\
    \noalign{\hrule height 1.0pt}
  \end{tabular}
}
\end{table*}

\begin{table*}[htb]
\centering{
\caption{Execution time divided by $L$ as a function of $L$ and the inner 
iteration number $j$ to solve the Wilson-Dirac equation with the
preconditioned block BiCGGR method. Results are averaged over 10
 configuration samples. Fail means how many configuration samples
out of ten show the convergence failure.
}
\label{tab:time}
\newcommand{\cc}[1]{\multicolumn{1}{c}{#1}}
   \begin{tabular}{r|r|r|r|r|r|r}
    \noalign{\hrule height 1.0pt}
    \multicolumn{7}{c}{$\kappa = 0.1359$}\\
    \multicolumn{1}{c}{$L$} & \multicolumn{1}{c}{$j=0$} & \multicolumn{1}{c}{$j=6$} & \multicolumn{1}{c}{$j=12$} & \multicolumn{1}{c}{$j=18$} & \multicolumn{1}{c}{$j=24$} & \multicolumn{1}{c}{$j=30$} \\
    \noalign{\hrule height 1.0pt}
     $1$ & $   1511.0$ & $   1593.3$ & $   1649.6$ & $   1727.5$ & $   1773.9$ & $   1812.2$ \\
    \hline
     $2$ & $    750.2$ & $    653.5$ & $    659.8$ & $    686.5$ & $    702.7$ & $    713.6$ \\
    \hline
     $4$ & Fail: $7$/$10$ & Fail: $1$/$10$ & $    330.3$ & $    329.1$ & $    338.3$ & $    346.4$ \\
    \hline
     $6$ & Fail: $10$/$10$ & $    230.3$ & $    229.2$ & $    230.0$ & $    236.2$ & $    237.8$ \\
    \hline
     $8$ & Fail: $10$/$10$ & Fail: $1$/$10$ & $    187.1$ & $    192.5$ & $    194.1$ & $    197.8$ \\
    \hline
     $10$ & Fail: $10$/$10$ & Fail: $1$/$10$ & $    167.2$ & $    167.7$ & $    172.4$ & $    174.0$ \\
    \hline
     $12$ & Fail: $10$/$10$ & $    168.7$ & $    150.8$ & $    154.1$ & $    159.6$ & Fail: $1$/$10$ \\
    \noalign{\hrule height 1.0pt}
  \end{tabular}
}
\end{table*}

\section{Conclusions and discussions}
\label{sec:conclusion} 

In this paper we present an evidence that the convergence behavior 
of the block BiCGGR can be improved by the preconditioning technique.
Our numerical tests show that the rank loss problem is remedied 
by the use of the inner solver with the Jacobi method 
as a preconditioner. As an optimized choice of $L$ and the inner
iteration $j$ we can achieve 90\% cost reduction in terms of
the execution time.

There remains a couple of future works.
Firstly, it is worthwhile to search a better preconditioner which 
assures the convergence for wider range of $L$ with less
computational cost.
Secondly, it is important to investigate why the
preconditioner allows us to avoid the rank loss problem. 
Thirdly, we plan to apply the
preconditioned block BiCGGR method to one of the state-of-the-art gauge
configurations generated by the PACS-CS Collaboration\cite{pacscs_nf3}.
Fourthly, it is interesting to make a direct comparison 
of the algorithmic efficiency between the preconditioned  
block BiCGGR method and other multiple right-hand-side 
methods\cite{luscher,orginos,wilcox,boston}.

\section*{Acknowledgments}
Numerical calculations for the present work have been carried out
on the T2K-Tsukuba computer at the University of Tsukuba.
This work is supported in part by Grants-in-Aid for Scientific Research
from the Ministry of Education, Culture, Sports, Science and Technology
(Nos.
20800009,   
18540250).    



\end{document}